\def\chapter#1{\vskip .5cm {\bf #1} \vskip .3cm}
\def\pl {\partial}
\def\D {\Delta}
\def\ps {\vskip .3cm}

\ps
\centerline{\bf A Real-Space Discrete Inverse Renormalization
Group Method.}
\ps
\centerline{Javier Rodr\'{\i}guez Laguna. {\sl IMAFF-CSIC Serrano 123, 
28006 Madrid.}}
\ps

Abstract. A numerical version of a real-space of
the Inverse Renormalization Group (IRG) proposed in [1] is developed. It
has been tested to obtain the scaling behavior of the random-forced
heat equation in the short scales limit. Prospectives are described,
and the most important target for the procedure is fully developed
turbulence.


\chapter{1. Introduction.}

Fully developed turbulence stands as one of the most important unsolved
problems in classical physics. The current description depends upon the
qualitative picture developed by Richardson and its quantitative counterpart
in Kolmogorov's work [2]. According to this view, fluctuations are settled
at the flow in a typical length scale $L$ due to a random stirring force
or topographic irregularities. These fluctuations create vortices at the
``integral scale'' $L$, which develop other vortices of smaller size. 
The process stands producing lesser and lesser vortices until the viscous 
scale $L_\eta$ is reached.
Then the ``fluctuating-energy cascade'' is stopped and viscous dissipation
takes place.
\ps

The main quantitative prediction of Kolmogorov theory is the value of the
scaling exponents for the velocity correlators
$<(v(x)-v(y))^n>\approx |x-y|^{n/3}$. The experimental data are not completely
in accordance with this prediction, just as mean-field theories are usually
not completely in agreement with experiment [3]. A Renormalization Group (RG)
analysis may enhance the results in a scale-invariant theory. Fully
developed turbulence, according to Richardson-Kolmogorov picture is
a scale-invariant theory, which suggests that a RG method might be 
implemented. 
\ps

As it is argued in [7], turbulence and field theory are
concerned with opposite limiting regimes. The integral scale is equivalent
to a cutoff, but the physically interesting scales are {\sl below} that
cutoff scale. Reference [1] puts forward this argument in order to explain 
the failure of usual RG procedures applied to turbulence phenomena. The
main problem is the appearance of an infinite number of expanding directions
in the interaction parameters space. It is argued by Gawedzki et al. that
a renormalization scheme fixing the integral scale (i.e.: the cutoff) and 
studying ever-decreasing scales would give a finite RG expansion. This
procedure has been named {\bf Inverse Renormalization Group}. The scaling
behaviour of Kraichnan's passive scalar model [5] via IRG is performed in
reference [1].
\ps

The motivation of this work is the development of a numerical
implementation of a real-space IRG method, which we shall name Discrete
Inverse Renormalization Group (DIRG).


\chapter {2. The Real-Space Discrete IRG method.}

The numerical procedure will be described in this section. Every numerical
calculation about a partial differential equation involves a discretization
of space and time. Our procedure adopts the common discretization of time
($t_n=n\D t$), but the space discretization is peculiar. Although we shall
restrict ourselves to evolution (parabolic) equations, this restriction
is not essential. Let us write the equation as

$$\pl_t u(t,x)={\cal O}u(t,x)$$

\noindent where ${\cal O}$ is any local operator, without any 
{\sl a priori} retrictions.
\ps

The calculations shall be particularized to a 2-dimensional square domain
with side length $L$
and the topology of a torus (periodic boundary conditions), but we should
remark that all these conditions might be easily removed.
The square is divided by a coarse grain mesh into $3\times 3$ little squares.
(See figure 1.)
The central little square is now divided into other $3\times 3$ lesser
squares, being the later grid 3 times finer. We may repeat the process
up to any desired depth level. Let us call $P$ the level. The
procedure described so far shall be referred to as a Discrete IRG
(DIRG).
\ps
 
The cells are indexed by three numbers: $C^n_{ij}$ denotes the cell
in the $i^{th}$ column and $j^{th}$ row of the $n^{th}$ level. (See
figure 2.) The indices run as follows: $i,j\in\{1\ldots3\}$ and
$n\in\{ 1\ldots P\}$.
The DIRG description of the $u(t,x)$ field consists of the $3\times 3
\times P$ values $u^n_{ij}(t)$ which represent the mean value of the
field in the correspondent cell: 

$$u^n_{ij}(t)=\int_{C^n_{ij}} dx u(t,x)$$

The $u^n_{ij}(t)$ must satisfy the consistency constraint: $u^n_{22}(t)=
{1\over 9} \sum_{ij} u^{n+1}_{ij}(t)$ for all time. Initial data
should be provided: the values of all the $u^n_{ij}(t_0)$ at a given
time $t_0$. 
\ps

The following step is to set up DIRG algebraic equations by means
of the suitable discretization of the partial derivatives
equation. The essential point is the discretization of the derivative
operators. Although there are other possible alternatives we have
chosen an scheme which we have named ``outward scheme'', that shall
be exemplified with the explicit calculation of the action of $\pl_x$  
and $\pl_y$ on $u^n_{12}(t)$:

$$(\pl_x u)^n_{12}(t)={1\over 2\cdot3^{n-1}L} [u^n_{13}(t)-u^n_{11}(t)]$$
$$(\pl_y u)^n_{12}(t)={1\over 3^{n-1}L} [u^n_{12}(t)-u^{n-1}_{12}(t)]$$

Derivative operators are taken, when possible, by comparing with $u$-values
in the same depth level, following the central scheme. When this is not
possible, as in the $\pl_y$ case in the former example, we compare the
$u$-value at the actual cell with $u$-values in the {\sl outer shell},
not in the inner one. This scheme is not the only one that is compatible
with the DIRG scheme, because it privileges the relation of the evolution
of the $u$-values at depth level $n$ with those at outer depth levels,
provoking an ``inward information-flow'', which is characteristic of
the Richardson-Kolmogorov view of turbulence.


\chapter {3. The DIRG for the random-forced heat equation.}

We have tested the DIRG method with an exactly soluble equation:
the heat equation with a random source term:

$$\pl_t T(t,x) = \kappa\nabla^2 T(t,x) + f(t,x)$$

\noindent where $f(t,x)$ is a gaussian random process, with zero
mean and covariance

$$<f(t,x)f(t',x')>=\delta(t-t')C(x-x')$$

\noindent with $C(x-x')$ falling quickly to zero at large distances and
nearly constant for $r\equiv |x-x'|<L$. 
\ps

The statistics of the solution field may be obtained by a Martin-Siggia-Rose 
formalism or otherwise [1,4]. If we let $t_0\to\infty$ then the mean value 
$<T(t,x)>=0$, and the two-point function depends on the covariance of
the random force:

$$\left< T(t_1,x_1) T(t_2,x_2) \right> =\int e^{|t_1-t_2|\kappa k^2 -
ik(x_1-x_2)} {\hat C(k)\over 2\kappa k^2} d\hat k $$

\noindent where $d\hat k={dk\over (2\pi)^{d}}$, as usually. Notice that
the Fourier transform of the covariance of the random force is now taking
the role of an ultraviolet cutoff. 
\ps

Now we may try to obtain the asymptotic behaviour of this correlation
function. We shall assume that the force acts only at large distances.
In other words, if we let $L$ be such a typical large length scale and
$\Lambda={1\over L}$,

$$\hat C(k)= C\cdot \theta(|k|-\Lambda)$$

The correlation function now gives ($r=|x_2-x_1|$, $\D t=t_2-t_1$):

$${C\over \kappa r}\int_0^\Lambda dk {\sin kr\over k} e^{\D t \kappa k^2}$$

The equal-time correlator becomes, making $x=r/L=\Lambda r$,

$$<T(x_1)T(x_2)>={C\over \kappa r} \int_0^{\Lambda r} dx {\sin x\over x}$$

For large distances it is easily seen that the correlation falls 
as $r^{-1}$, but the purpose of this calculation is to obtain the 
short-distance (compared to $L$) behaviour of the correlation function.
The integrand should be approximated to second order
in order to obtain:

$$<T(x_1)T(x_2)>\approx {C\Lambda\over \kappa} - {C\Lambda^3\over 12} r^2 +
{\cal O}(r^4)$$

This means that, substracted the constant, we obtain a scaling exponent
2, when looking at short distances. This exponent is also obtained in
our DIRG approach. Let us explain how is this result achieved.
\ps

The DIRG analysis of this equation follows the steps of the preceding
section. A random $f(t,x)$, decorrelated in time, is introduced only at
the outmost level. All depth levels $n>1$ evolve with a free heat equation,
thus losing very early in the calculation all information referring to
the initial condition. Fluctuations ``cascade'' inwards and induce a
correlation function $T(t,x)T(t,x+\D x)$ which is time-averaged separately
at each depth level. 
\ps

Figure 3 plots the logarithm of this quantity 
against depth level. 
This is equivalent to a log-log graph of the correlation function with
the space axis inverted. 
It is clearly seen that the resulting curve approaches
asymptotically a straight line with slope $2.06 \pm 0.09$.
The analytical result, which gives a behaviour
of $A+Br^2$, is thus satisfied by our approximate DIRG method.
\ps

This model is not completely suitable for an DIRG study due to the 
dimensionful constant $\kappa$, which prevents scale-invariance. As
a consequence, it is difficult to reach very high depth levels. In
our calculations we have only simulated 9 levels, but they are enough
to appreciate the scaling behaviour (figure 3 shows only 8 levels 
because the last one is used in order to determine the other constant).


\chapter {4. Conclussion and prospectives.}

The main results expected in any DIRG analysis are scaling exponents of
certain functions, specially correlation functions (type $<\prod_i^N T_i
(t_i,x_i)>$), structure functions (type $<(T(t,x)-T(t',x'))^n>$). 
Generally speaking, all statistical parameters having scaling properties,
or even multifractal behaviour, are possible candidates. Exotic scaling
behaviours, such as discrete scale invariance [6], might profit from
DIRG tools.
\ps                                                               

Notice that one of the advantages of the DIRG method over direct numerical
simulation is that it is able to capture the scaling behaviour over
various scale-decades saving a great deal of computer power. In order to
simulate a given factor $S\equiv L/L_\eta$ usual numerical
calculations require ${\cal O}(S^d)$, being $d$ the space dimension,
while DIRG needs only ${\cal O}(d\ \log S)$. The key is that DIRG focus
on scaling behaviour, but it is not trying to obtain the whole dynamics.
\ps

The DIRG scheme described so far appears to be specially well suited for
dealing with turbulence problems, according to the Richardson-Kolmogorov
picture. The author is working on an implementation of the DIRG
to Kraichnan's passive-scalar problem. This is an exactly soluble model [5]
in which we suppose certain time-decorrelated random velocity field and
a contaminant (such as a solute or even temperature in the adiabatic limit)
is advected through it. The exact solution shows a scaling universal
behaviour with anomalous scaling exponents, i.e. not the ones given by
naive dimensional analysis.
\ps

The main objective of these calculations are fully developed turbulence,
in order to obtain numerically, following Gawedzki's ideas, the statistical
properties of the flow. There is no point in remarking the importance for
diverse branches of applied physics, such as astrophysics, geophysics,
aeronautical engeneering, meteorology... of a first-principles derivation
of the scaling behaviour of the solution of any ``turbulent equation'':
Navier-Stokes, Magneto-hydrodynamics...


\chapter {Acknowledgements.}

The author wishes to acknowledge Germ\'an Sierra and Silvia N. Santalla for
useful discussions. Silvia N. Santalla is also acknowledged for her
collaboration in the preparation of the postscript figures.


\chapter {Bibliography:}

[1] K. Gawedzki {\sl Turbulence under a magnifying glass.} Lectures
given at the 1996 Carg\`ese Summer Institute. Available at chao-dyn/9610003.

D. Bernard, K. Gawedzki, T. Hurd and A. Kupiainen {\sl Inverse 
renormalization group.} In preparation. 

[2] V.S. L'vov {\sl Scale invariant theory of fully developed hydrodynamic
turbulence -- Hamiltonian approach.} Phys. Rep. {\bf 207} (1991) 1.

A.N. Kolmogorov {\sl The local structure of turbulence in incompressible
viscous fluid for very high Reynolds number.} Reprinted at Proc. Roy. Soc.
Lond. {\bf A 434} (1991) 9.

[3] R.H. Kraichnan {\sl Turbulent cascade and intermittency growth.}
Proc. Roy. Soc. Lond. {\bf A 434} (1991) 65.

[4] P.C. Martin, E.D. Siggia and H.A. Rose {\sl Statistical dynamics
of classical systems.} Phys. Rev. {\bf A 8} (1973) 423.

[5] R.H. Kraichnan {\sl Anomalous scaling of a randomly advected passive
scalar.} Phys. Rev. Lett. {\bf 72} (1994) 1016.

D. Bernard, K. Gawedzki and A. Kupiainen {\sl Anomalous scaling in the
N-point functions of passive scalar.} Phys. Rev. {\bf E 54} (1996) 2564.

D. Bernard, K. Gawedzki and A. Kupiainen {\sl Slow modes in passive
advection.} J. Stat. Phys. {\bf 90} (1998) 519. 

[6] D. Sornette {\sl Discrete-scale invariance and complex dimensions.}
Phys. Rep. {\bf 297} (1998) 239.

[7] G. Eyink and N. Goldenfeld {\sl Analogies between scaling in turbulence,
field theory and critical phenomena.} Phys. Rev. {\bf E 50} (1994) 4679.

\vfill
\eject
 

\chapter {Figure Captions.}

Figure 1. The DIRG space discretization. The initial square domain is
divided into a $3\times 3$ mesh, which is itself divided into lesser and
lesser squares, just as in Richardson-Kolmogorov picture.
\ps

Figure 2. The nomenclature of the cells in the DIRG space discretization.
Notice also the lines between cells. Distances are taken along them in
order to obtain the derivative operators.
\ps

Figure 3. DIRG results for the random forced heat equation. The vertical
axis represents the logarithm of the two-point function and the horizontal
axis is the depth level. A $-2$ slope straight line is superimposed to
see the accordance for small scales (high depth level).

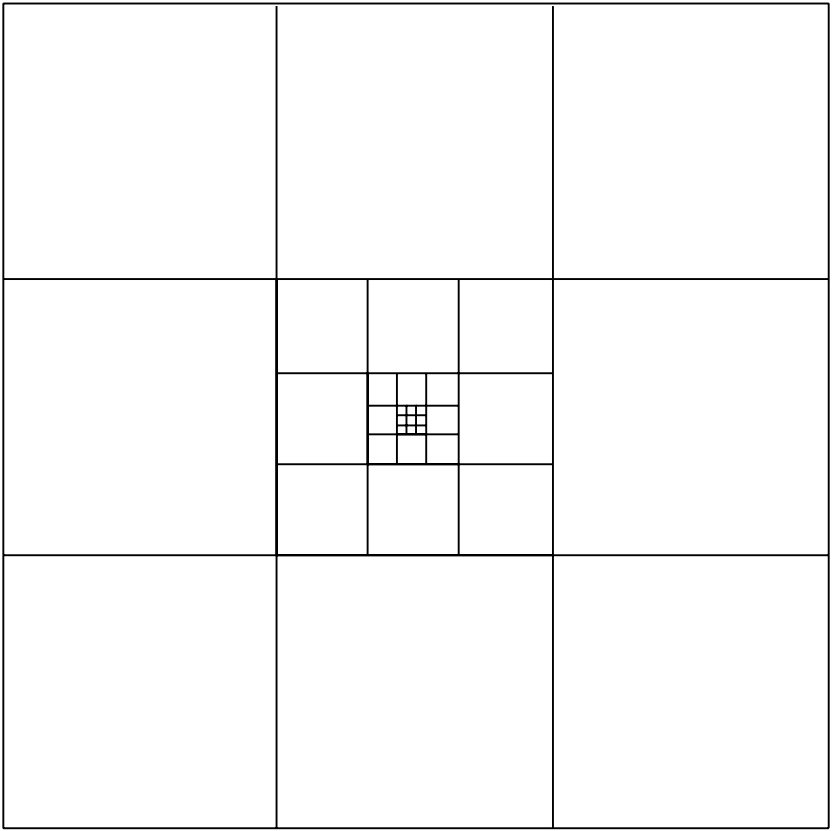
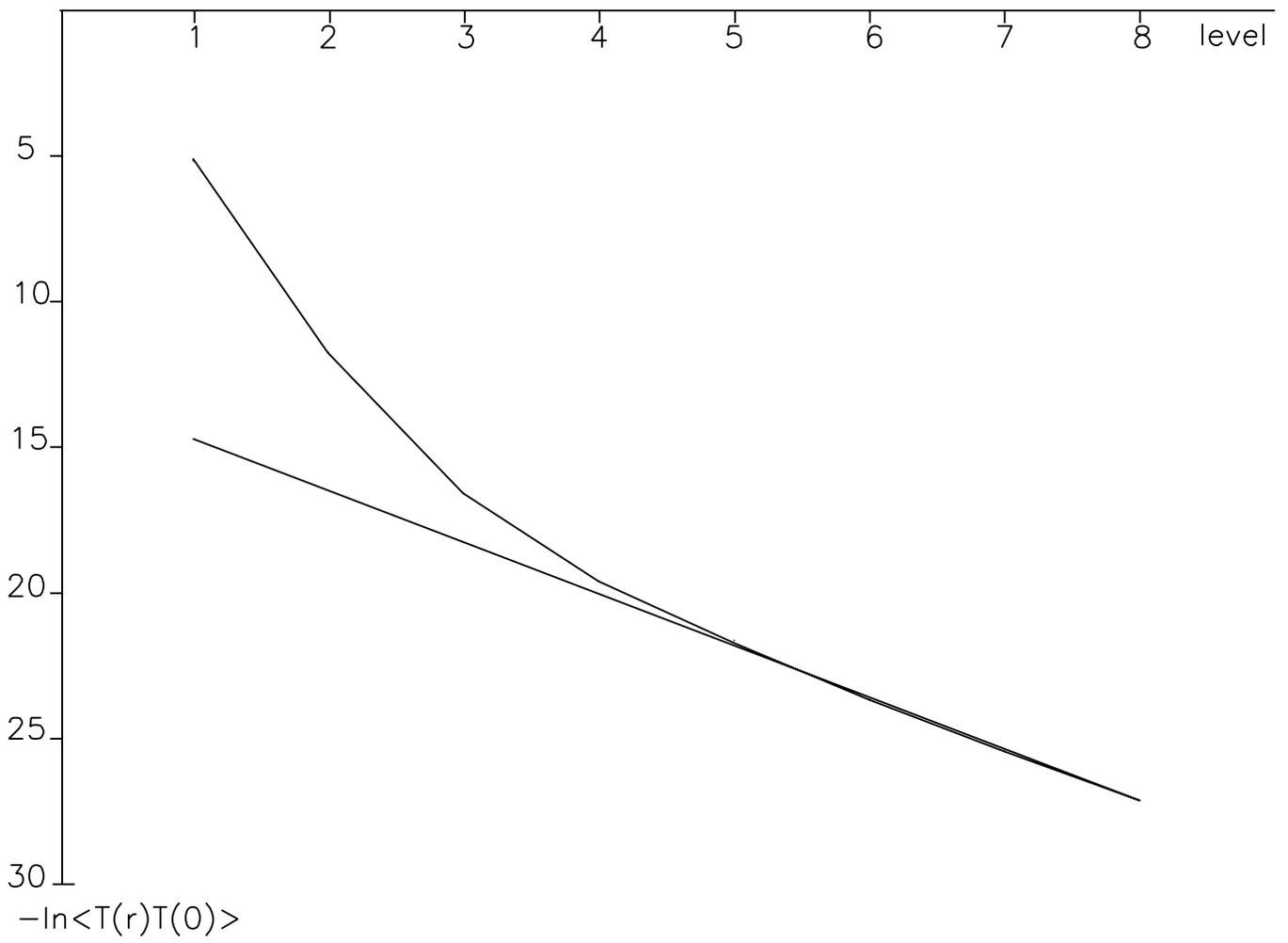
\bye